\documentclass[fleqn,usenatbib]{mnras}

\usepackage{graphicx}	% Including figure files
\usepackage{amsmath}	% Advanced maths commands
\usepackage{amssymb}	% Extra maths symbols
\usepackage{xcolor}

\newcommand{\dw}{Sahlmann\,2\,AB}
\newcommand{\dwa}{Sahlmann\,2\,A}
\newcommand{\dwb}{Sahlmann\,2\,B}
\newcommand{\dwr}{Reid\,1\,AB}
\newcommand{\dwra}{Reid\,1\,A}
\newcommand{\dwrb}{Reid\,1\,B}

\title[ \dw]{New constraints on the minimum mass for thermonuclear lithium burning in brown dwarfs 
\thanks{\textit{Based on observations made with the Gran Telescopio Canarias (GTC), installed in the Spanish Observatorio del Roque de los Muchachos of the Instituto de Astrof\'isica de Canarias, in the island of La Palma (programmes GTC100-20B and GTC51-21B).}} }

\author[Mart\'{i}n, Lodieu \& del Burgo]{
E.\ L.\ Mart\'{i}n,$^{1,2,3}$ \thanks{E-mail: ege@iac.es (EM)}
N.\ Lodieu,$^{1,2}$ 
C.\ del Burgo,$^{4}$ 
\\
$^{1}$Instituto de Astrof\'isica de Canarias (IAC), Calle V\'ia 
L\'actea s/n, E-38200 La Laguna, Tenerife, Spain\\
$^{2}$Departamento de Astrof\'isica, Universidad de La Laguna (ULL), 
E-38206 La Laguna, Tenerife, Spain\\
$^{3}$Consejo Superior de Investigaciones Cient\'ificas, E-28006 Madrid, Spain\\
$^{4}$Instituto Nacional de Astrof\'{\i}sica, \'Optica y Electr\'onica, Luis Enrique Erro 1, Sta. Ma. Tonantzintla, Puebla, Mexico\\
}

\date{Accepted 2021 10 08. Received 2021 06 04; in original form \today{}}
\pubyear{2021}

\begin{document}
\label{firstpage}
\pagerange{\pageref{firstpage}--\pageref{lastpage}}
\maketitle

\begin{abstract}
 The theory of substellar evolution predicts that there is a sharp mass boundary between lithium and non-lithium brown dwarfs, not far below the substellar-mass limit. The imprint of thermonuclear burning is carved on the surface lithium abundance of substellar-mass objects during the first few hundred million years of their evolution, leading to a sharp boundary between lithium and non-lithium brown dwarfs, so-called, the lithium test.  
The theoretical predictions can be tested by comparing with observations of lithium in the individual components of binaries with dynamical masses measured from orbital motions.  New optical spectroscopic observations of the binaries  \href{http://simbad.u-strasbg.fr/simbad/sim-basic?Ident=DENIS+J063001.4-184014&submit=SIMBAD+search}{DENIS\,J063001.4$-$184014AB} and \href{http://simbad.u-strasbg.fr/simbad/sim-basic?Ident=DENIS+J225210.7-173013&submit=SIMBAD+search}{DENIS\,J225210.7$-$173013AB} obtained using the 10.4-m Gran Telescopio de Canarias are reported here. They allow us to re-determine their combined optical spectral types (M9.5 and L6.5, respectively) and to search for the presence of the Li\,{\small{I}} resonance doublet. The non detection of the Li\,{\small{I}} feature in the combined spectrum of DENIS\,J063001.4$-$184014AB is converted into estimates for the depletion of lithium in the individual components of this binary system.  In DENIS\,J225210.7$-$173013AB we report the detection of a weak Li\,{\small{I}} feature which we tentatively ascribe as arising from the contribution of the T3.5-type secondary.  
Combining our results with data for seven other brown dwarf binaries in the literature treated in a self-consistent way, we confirm that there is indeed a sharp transition in mass for lithium depletion in brown dwarfs, as expected from theoretical calculations. We estimate such mass boundary is observationally located at  51.48$^{+0.22}_{-4.00}$ $M_\mathrm{Jup}$, which is lower than the theoretical determinations.  
\end{abstract}


\section{Introduction}
Brown dwarfs are objects that do not have enough mass to settle on the H-burning main-sequence because they develop cores dominated by degenerate electrons as they keep contracting and cooling forever. 
For solar chemical composition, the limiting mass of a star was found to be located at 73 $M_\mathrm{Jup}$ for completely convective numerical models \citep[]{Kumar:1963zr}, which is close to the modern theoretical value of 78.5 $M_\mathrm{Jup}$ \citep[e.g.,][]{Baraffe:2015aa}.   

Since lithium is destroyed by proton bombardment at temperatures around 2$\times$10$^6$ K, just below those required for H-fusion, there is a minimum mass for thermonuclear lithium burning (MMLB), which is slightly below the stellar/substellar boundary. The MMLB was estimated to lie at around 63 $M_\mathrm{Jup}$\citep[]{1991MmSAI..62..171Pozio, 1993ApJ...413..364Nelson}, and the search for the Li\,{\small{I}} resonance doublet in optical spectra of brown dwarf candidates became known as the lithium test \citep[]{Magazzu:1993kx}. 
Objects cooler than effective temperature 2700 K, corresponding to spectral type M7 and later  \citep[]{Rajpurohit:2013fk},  
and with lithium detected, are considered as certified bona fide brown dwarfs with masses below 63 $M_\mathrm{Jup}$ \citep[e.g.,][]{1998MNRAS.296L..42Tinney, 2000AJ....120..447Kirkpatrick}.

Not long after the first bona fide brown dwarf was discovered in the Pleiades cluster \citep[]{1995Natur.377..129R}, the first brown dwarf binary was revealed \citep[]{Basri:1999kx}. 
Careful astrometric and spectroscopic monitoring of the gravitational interactions between the components of ultracool binary systems are able to yield dynamical masses that lie below the minimum stellar mass, supporting the brown dwarf status of one or both components \citep[e.g.,][]{Zapatero-Osorio:2004fu, Burgasser:2009fk, Dupuy:2009lk}. 
In some cases the brown dwarf status is confirmed both by dynamical mass and by the detection of lithium \citep[e.g.][]{Sahlmann:2015ab,Dupuy:2017aa}.  

A long-term astrometric monitoring campaign using the FORS2 optical camera mounted on the Very Large Telescope (VLT) focused on 20 ultracool dwarfs \citep{Sahlmann:2014aa} that had been discovered in a photometric and proper motion study of 4800 squared degrees at low galactic latitude covered by the DENIS survey \citep{Phan-Bao:2008fr}. 
Two of those objects were found to be binaries. Here we focus on 
DENIS J063001.4$-$184014AB (hereafter \dw{}, inspired by the Washington Double Star catalog convention of naming objects after the discoverer of the binary nature), which was classified as an M8.5 dwarf from optical low-resolution reconnaissance spectroscopy and it was found to be an astrometric binary with an orbital period of $3.1\,$years \citep{Sahlmann:2015ac}. The dynamical mass of each component in the binary has been obtained from combining the astrometric data with spatially resolved near-infrared imaging, and radial velocities for the primary \citep{2021MNRAS.500.5453S}.  
These authors noted that there was no sign of strong Li\,{\small{I}} resonance doublet in a high-resolution optical spectrum of the system, despite the low masses of $M_1 = 54^{+10}_{-9}$~$M_\mathrm{Jup}$ and $M_2 = 54^{+6}_{-5}$~$M_\mathrm{Jup}$ of the components, but they could only place a upper limit of 1\,\AA{} on the measurable pseudo-equivalent width (pEW) because of low signal to noise ratio. 

 The second focus of this work is on the binary DENIS\,J225210.7$-$173013AB (hereafter \dwr{}). This object was identified as an L7.5 nearby ultracool dwarf from near-infrared spectroscopic follow-up of candidates detected in the DENIS survey using CGS4 on UKIRT  \citep{kendall04}. High resolution images obtained with the Near-Infrared Camera and Multi-Object Spectrometer (NICMOS) on the Hubble Space Telescope (HST) resolved the object into two components with an angular separation of 0.13 arcsec \citep{reid06}. These authors also obtained low-resolution near-infrared spectra with Spex on the NASA IRTF telescope and estimated individual spectral types of L6$\pm$1 and T2$\pm$1 for \dwra{} and \dwrb{}, respectively. Monitoring of the orbital motion of the \dwr{} system was carried out with the Canada-France-Hawaii Telescope, the Keck telescopes and HST \citep{Dupuy:2017aa}. These authors derived an orbital period of $8.8\,$years, and estimated a spectral type of L4$\pm$1 and a mass of $M_1 = 59^{+5}_{-5}$~$M_\mathrm{Jup}$ for the primary, and a spectral type of T3.5$\pm$0.5 and a mass of $M_2 = 41^{+4}_{-4}$~$M_\mathrm{Jup}$ for the secondary. No detection of the Li\,{\small{I}} resonance doublet was reported by \citet{2008AJ....136.1290Reid} in the optical spectrum of the integrated light from the binary.    

This paper is structured as follows: In Section 2, we present new optical spectra specifically aimed at searching for lithium in these two binaries. It turns out that we do not detect lithium in \dw{}, which allows us to infer that both components have experienced Li depletion, and we do detect weak lithium in \dwr{} that we interpret as likely coming from the secondary. In Section 3, our results are put in context with other very low-mass (VLM) binaries for which dynamical masses and lithium data are available in the literature. The consistency between ages and metallicities obtained from stars and brown dwarfs for three VLM binaries that are members in hierarchical multiple systems is discussed in Section 4. In Section 5, we compare our results with calculations of lithium burning, and we confirm that there is a sharp transition in dynamical mass between non lithium bearing and lithium bearing brown dwarfs, as expected from the theory of substellar evolution. However, the location of the MMLB from the observational side is not completely consistent with theoretical calculations. Section 6 provides a summary of our conclusions and a discussion of future prospects that are opened by this research.   

\section{Observations, data reduction and spectroscopic measurements}

 Optical spectra of the integrated light of the \dw{} and the \dwr{} systems were collected in service mode with the Optical System for Imaging and low Resolution Integrated Spectroscopy
\citep[OSIRIS;][]{cepa00} instrument on the 10.4-m Gran Telescopio de Canarias (GTC). 
\dw{} was observed on 2021 February 18th (UT) for programme GTC100-20B, 
%(PI E.L. Mart\'\i n)
and \dwr{} was observed on 2021 August 4th (UT) for programme GTC51-21A. In the later programme, a spectrum of the known \citep{Cruz:2003aa} L1-type lithium dwarf 2MASS\,J20575409$-$0252302 (hereafter 2M2057-02) was taken with the same instrumental configuration. 
The Principal Investigator (PI) for both programmes was the lead author of this paper. 

The R1000R grating covering the 5100--10000\,\AA{} wavelength range and a slit width of 1.0 arcsec, yielding a spectral resolving power of 672 at the central wavelength, was used. The sky conditions were clear and with subarcsecond seeing. Four on-source integrations with individual exposure times of 630 s were obtained for \dw{}, six exposures of 1500 s for \dwr{} and also six exposures of 1260 s each for 2M2057+0252. The targets were nodded by 15 arcsec on the slit to allow us for sky subtraction. Parallactic angle was selected to minimize atmospheric refraction. Flat field lamp, arc lamp, and dark frame exposures were used for standard calibrations within the IRAF environment
\citep{tody93}. 

\begin{figure}
\centering
\includegraphics[width=\linewidth]{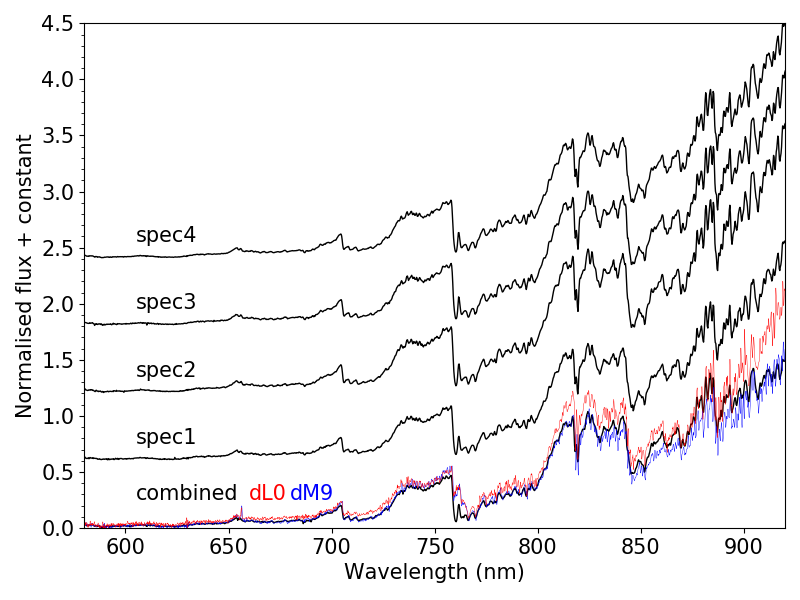}
\includegraphics[width=\linewidth]{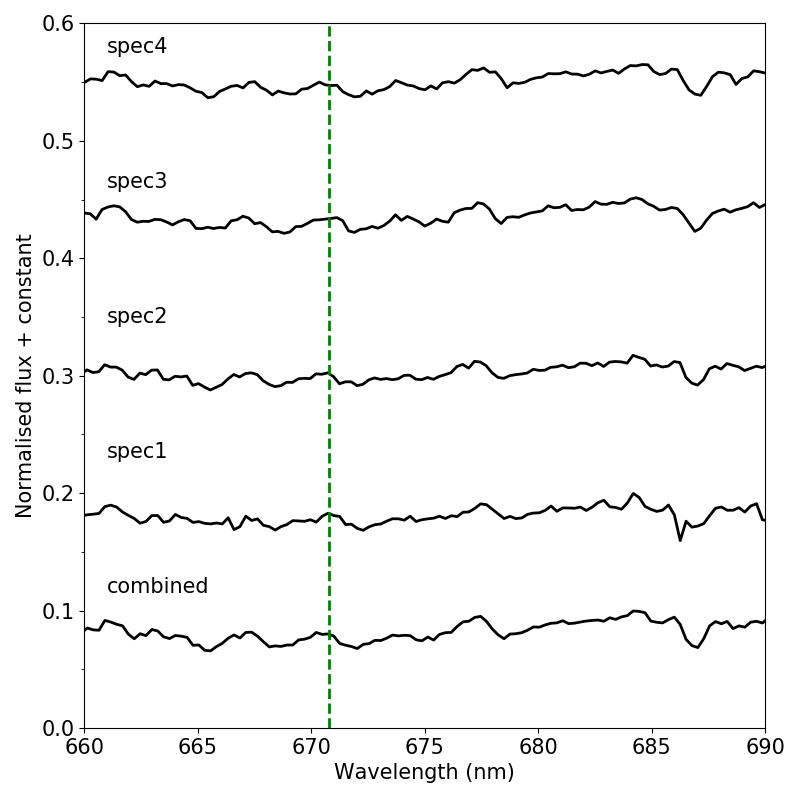}
\caption{Full OSIRIS spectra (top panel) of \dw{} compared with standard ultracool dwarfs of spectral type M9 (blue line) and L0 (red line) from the Sloan archive, and zoom around the Li\,{\small{I}}
resonance doublet (bottom panel). The position of the Li\,{\small{I}} feature at 670.8 nm is marked with a green horizontal dashed line.}
\label{fig:midres}
\end{figure}

\begin{figure*}
\centering
\includegraphics[width=\linewidth]{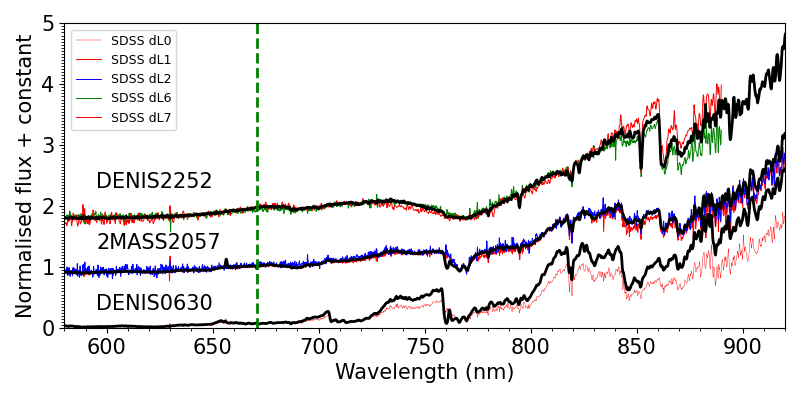}
\caption{Full OSIRIS spectra of DENIS0603 (\dw{}), 2MASS2057 and DENIS2252 (\dwr{}) compared with standard ultracool dwarfs of spectral types between dL0 and dL7 from the Sloan archive.}
\label{fig:midres2}
\end{figure*}

The four individual spectra of \dw{} and the mean combined one are shown in Fig.\ \ref{fig:midres}.  The mean combined spectrum of \dw{}, \dwr{} and the L1-type lithium dwarf 2M2057-02 are displayed in Fig.\ \ref{fig:midres2}. 

Following the same procedure as in our previous work in the nearby open clusters the Hyades \citep{Martin:2018aa} and Coma Berenices \citep{martin20a},  
where we used similar instrumental setup, spectral types were derived by comparison with field late-M and L dwarfs from the Sloan database \citep{bochanski10, Schmidt:2010aa}. 
The GTC/OSIRIS mean spectrum of the \dw{} system was found to be intermediate between the M9 and the L0 standards, and hence it is classified as M9.5$\pm$0.5 type, i.e., one spectral subclass later than the classification of M8.5 reported by \citet{Phan-Bao:2008fr}. 
 The GTC/OSIRIS mean spectrum of the \dwr{} system was found to be intermediate between the L6 and the L7 standards, and hence it is classified as L6.5$\pm$0.5 type, i.e., one spectral subclass earlier than it was estimated in \citet{Knapp:2004aa}. 

The following atomic spectral features were clearly detected in the spectra: H$_\alpha$ in emission, the Rb\,{\small{I}} resonance doublet at 780.03 and 794.76\,nm, the Na\,{\small{I}} subordinate doublet at 819.0\,nm and the Cs\,{\small{I}} resonance doublet at 852.11 and 894.30\,nm. Pseudo equivalent widths (pEW) were measured by direct integration of the line profiles with respect to the local pseudo-continuum using the task splot in the IRAF environment. Gaussian fitting was applied to each spectral line to help guiding the choice of the integration boundaries for pEW determination. 

\begin{figure}
\centering
\includegraphics[width=\linewidth]{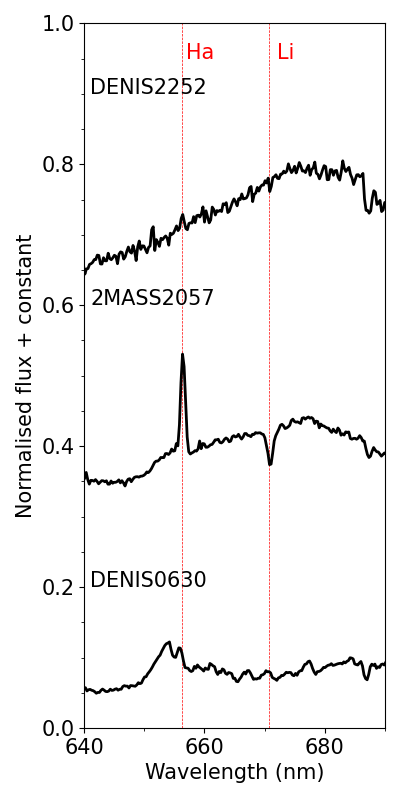}
\caption{Zoom around H$_\alpha$ and the Li\,{\small{I}}
resonance doublet (bottom panel) for the three targets observed in this work. The position of H$_\alpha$ and the Li\,{\small{I}} feature at 670.8 nm is marked with a red horizontal dashed line.}
\label{fig:midres3}
\end{figure}

The Li\,{\small{I}} feature was not detected in any of our GTC/OSIRIS spectra of \dw{}. A zoom on the Li\,{\small{I}} spectral region of each individual spectrum is shown in the lower panel of Fig.\ \ref{fig:midres}. Conversely, a weak Li\,{\small{I}} feature was detected in our mean combined GTC/OSIRIS spectrum of \dwr{} and a strong feature was detected in the L1 standard. These spectra are displayed in Fig.\ \ref{fig:midres3}.

Signal-to-noise ratios (SNR) in a spectral region relatively free from strong absorption features in the GTC spectra of our targets (between 680.5 and 683.5 nm) were measured with key ''m'' in the splot task to estimate the sensitivity of the data to detection of the Li\,{\small{I}} resonance doublet at 670.8 nm. 
 
We used a spectral region slightly different than that adopted by \citet{Martin:2018aa} because we could see an obvious absorption bandhead in the spectrum of \dw{} in the region between 675.0 and 680.0 nm. Our SNR estimates are conservative because there is no region completely free from absorption features in the spectra of ultracool dwarfs. 

The spectroscopic measurements and their uncertainties are summarized in Table \ref{tab_Phan2:pew}.  
The adopted 1$\sigma$ upper limits and the uncertainties on the pEW measurements come from a conservative assessment of the errors obtained from the SNR of the combined spectra using the equation discussed in \citet{Martin:2018aa}, and the errors obtained manually using the task splot with the options for direct integration and gaussian fitting. 
The weak H$_\alpha$ in emission and the line strength of the Cs\,{\small{I}} and Na\,{\small{I}} measured in \dw{} are normal for an old field M9.5 dwarf of solar-metallicity. There is no spectroscopic sign of youth in the Na\,{\small{I}} doublet of the \dw{} system which is an indicator of low surface gravity if it is weaker than the average for field dwarfs \citep{2010A&A...517A..53M}. The upper limit on the presence of the Li\,{\small{I}} in \dw{} is rather conservative because in fact we could not see any feature whose pseudo-equivalent width is pEW\,$\geq$\,0.1\,\AA{} around 670.8 nm in any of the spectra. 
Our adopted upper limit of 0.17\,\AA{} on the pEW of the Li\,{\small{I}} resonance doublet in \dw{} improves the sensitivity to lithium detection in \dw{} by a factor of almost six times over the upper limit of 1\,\AA{} reported in \citet{2021MNRAS.500.5453S} from a noisy (SNR\,$<$\,3) high-resolution UVES spectrum obtained with the European Southern Observatory Very Large Telescope. 

For \dwr{}, we also detect weak H$_\alpha$ in emission, which is unusual for a late-L dwarf because this chromospheric indicator is detected in less than about 20\% of field L5--L8 dwarfs \citep{pineda16}. The strength of the Cs\,{\small{I}} and Rb\,{\small{I}} features in \dwr{} are similar to those reported in the literature \citep{Martin99, 2000AJ....120..447Kirkpatrick, 2015A&A...581A..73Lodieu, pineda16}. The detection of a weak Li\,{\small{I}} feature is highly unusual for a late-L dwarf. 

\begin{table*}
\centering
\caption{Pseudo EWs of atomic lines in \AA{} estimated from the combined GTC/OSIRIS spectra of each target.
}
 \begin{tabular}{@{\hspace{0mm}}l c c c c c c c c @{\hspace{0mm}}}
 \hline
 \hline
Name  &  H$_\alpha$ & Li\,{\small{I}}  &  Rb\,{\small{I}}  &  Rb\,{\small{I}}  & Na\,{\small{I}}  & Cs\,{\small{I}} & Cs\,{\small{I}}  & SNR   \cr
 \hline
      & 656.28\,nm{} & 670.78\,nm{} & 780.03\,nm{} & 794.76\,nm{} & 819.0\,nm{} & 852.11 nm{} & 8943.0\,nm{} & 680.5--683.5 nm \cr
 \dwr{}     &  $-$0.9$\pm$0.6 &  0.5$\pm$0.1 &  4.9$\pm$0.1 &  5.6$\pm$0.1 & 4.9$\pm$0.6 &  5.1$\pm$0.1 &  4.5$\pm$0.1 &  26 \cr
 \dw{}              & $-$1.1$\pm$0.4 & $<$0.17     & 1.0$\pm$0.2 & 2.3$\pm$0.1 & 8.3$\pm$0.5 & 1.4$\pm$0.1 & 1.5$\pm$0.1 & 53 \cr
 2MASS2057-02 &  $-$12.6$\pm$0.2 &  4.5$\pm$0.1 &  2.4$\pm$0.1 &  2.5$\pm$0.1 &  5.6$\pm$0.5 &  1.6$\pm$0.1 &  1.9$\pm$0.1 &  35 \cr
\hline
\label{tab_Phan2:pew}
\end{tabular}
\end{table*}
 
\section{Dynamical mass and lithium in the individual components of VLM binaries}

\subsection{Lithium depletion in each component of Sahlmann 2 AB} 

To convert the adopted upper limit pEW Li\,{\small{I}} of $<$\,0.17\,\AA{} for the combined light of the \dw{} system into lithium detection limits for each component, the individual contributions to the flux around the spectral region of the Li\,{\small{I}} resonance doublet are needed. \citet{2021MNRAS.500.5453S} estimated $\Delta I_\mathrm{FORS2}$\,=\,2.8$\pm$0.5 mag from a Keck adaptive optics assisted spatially resolved imaging observation that gave $\Delta K_\mathrm{NIRC2}$\,=\,1.74$\pm$0.06 mag.    
These authors also carried out binary fitting of the low-resolution near-infrared spectral data of \dw{} from \citet{2019ApJ...883..205B} and concluded that the best-fit was M8.5$\pm$0.5 for \dwa{} and  L4.5$\pm$2.5 for \dwb{}. In this work those individual spectral types are adopted. Using as photometric reference the bright ($m_{R}$\,=\,15.051$\pm$0.014 mag; \citet{2005AJ....130..337C}) 
and nearby (d\,=\,$4.04\pm0.01$\,pc) M8.5 dwarf DENIS\,J104814.6$-$395606, a color of $R-K_{S}$\,=\,6.60$\pm$0.02 mag is assumed for \dwa{}. On the other hand, we take an average color of $R-K_{S}$\,=\,7.6$\pm$0.2 mag for the two L5 standards of \citet{2006PASP..118..659L}  
as representative of \dwb{}. From those colors, we obtain $\Delta R$\,=\,2.9$\pm$0.2 mag or flux ratio of 14.6$\pm$1.0 around the spectral region of the Li\,{\small{I}} resonance doublet. 

The second step is to estimate the upper limits on the pEW Li\,{\small{I}} for each component. The contribution to the flux of each component was accounted for by multiplying the pEW by a factor $1 + r$, where $r\,=\,Fratio\,\times\,Aratio$, where $Fratio$ is the ratio of the fluxes, and $Aratio$ the ratio of the areas \citep{1993AA...274..274M}. Assuming that both components in the \dw{} system have the same radii, the upper limit for \dwa{} is pEW Li\,{\small{I}} $<$\,0.18\,{\AA} and for \dwb{} is pEW Li\,{\small{I}} $<$\,2.7\,\AA{}. 

\begin{figure}
\centering
\includegraphics[width=\linewidth]{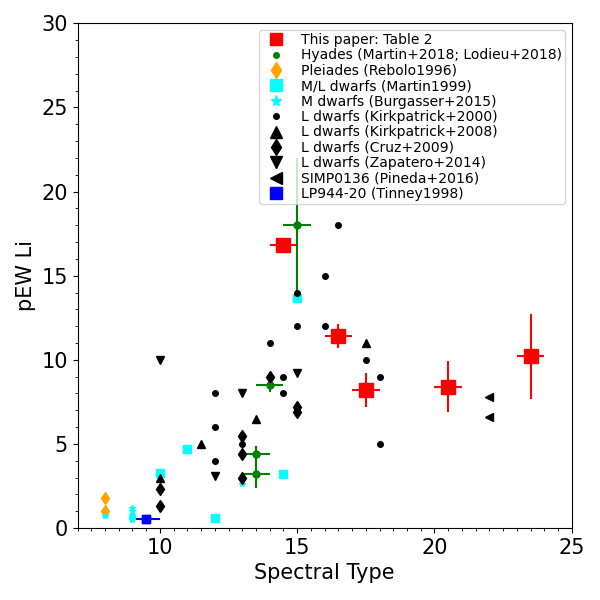}
\caption{Pseudo-equivalent widths (pEW) of Li\,{\small{I}} versus spectral type (L0 = 10; L5 = 15; T0 = 20; T5 = 25 in the abscissa axis) for ultracool dwarfs in the field and in open clusters. The following data sources have been used: \citet{Burgasser:2015aa} for field late-M and L dwarfs; \citet{Cruz:2009fk} for field L dwarfs; 
\citet{2000AJ....120..447Kirkpatrick} and 
\citet{Kirkpatrick:2008uk} for field L dwarfs; \citet{2015A&A...581A..73L} for the components of Luhman16AB; 
\citet{Martin99} for field late-M and L dwarfs; \citet{2018AA...615L..12Lodieu}  
and \citet{Martin:2018aa} for L dwarfs in the Hyades;  \citet{pineda16} for the T2 dwarf SIMP0136;  \citet{1996ApJ...469L..53Rebolo} for late-M brown dwarfs in the Pleiades; 
\citet{1998MNRAS.296L..42Tinney} for the M9 lithium brown dwarf LP 944-20; and \citet{Zapatero-Osorio:2004fu} for field L dwarfs. Note that the range of spectral types of brown dwarfs with Li\,{\small{I}} detections is between M8.5$\pm$0.5 and T3.5$\pm$0.5\@, which is the domain that we adopt for this study. }
\label{fig:sptLi}
\end{figure}

The third step is to compare the upper limit on the pEW Li\,{\small{I}} for each component to the upper envelope of pEW Li\,{\small{I}} in ultracool dwarfs as a function of spectral class. The pEW Li\,{\small{I}} depends strongly on the spectral type as illustrated in Fig.\ \ref{fig:sptLi}. Both components in the \dw{} system have about a factor of at least four weaker pEW Li\,{\small{I}} than the maximum values for their Li bearing counterparts of the same spectral type implying more than a factor of 10 Li depletion according to the calculations provided in \citet{Martin:2018aa}, i.e., they have destroyed more than 90\% of their initial Li content. Their Li depletion factors are estimated in more detail in Section 5. 

\subsection{Lithium in each component of Reid 1 AB}

We follow the same approach to disentangle the contributions to the lithium detection in the components of \dwr{}. Following \citet{Dupuy:2012fk}, we consider that the primary has a spectral type of L4.5$\pm$1.5, the secondary has a spectral type of T3.5$\pm$0.5, and the magnitude difference between them is 0.98$\pm$0.03 mag in the HST F110W-band.  Using the flux ratio at $R$-band between L4.5 and L8.5 derived from the $R-J$ data of \citet{2006PASP..118..659L}, a flux ratio of 3.1$\pm$0.6 in the $R$-band is obtained. This ratio is multiplied by a factor of 6.25 obtained from the flux ratio between the L8 and the T3 dwarfs illustrated in Figure 16 of \citet{Burgasser:2003} 
from Keck LRIS optical spectra to yield a flux ratio of 19.4$\pm$0.8. No lithium detection was reported by \citet{2008AJ....136.1290Reid} from optical spectra obtained with GMOS with resolving power of R\,=\,1220\@. They did note lithium detections with GMOS in eight L dwarfs and provided pEW Li\,{\small{I}} measurements between 3.5 and 12.2\,\AA{} for them. Our GTC spectra improve the sensitivity to Li\,{\small{I}} detection by almost an order of magnitude given that we are able to measure a pEW Li\,{\small{I}} $=$ 0.5 $\pm$ 0.1\,\AA{}. If this Li\,{\small{I}} feature comes from the primary, it would mean a very large Li depletion because it is much weaker than those measured in late-L dwarfs. On the other hand, if the Li\,{\small{I}} feature comes from the secondary, it implies an upper limit of pEW Li\,{\small{I}}\,$<$\,0.51\,\AA{} for \dwra{}, and pEW Li\,{\small{I}}\,$=$\,10.2$\pm$2.5\,\AA{} for \dwrb{}. Such a value of pEW Li\,{\small{I}} for \dwrb{} is similar to those reported for the T0.5 dwarf Luhman16B \citep{2015A&A...581A..73L} and the T2 dwarf SIMP0136+0933 \citep{pineda16}, which supports our interpretation that the Li\,{\small{I}} feature in our integrated light GTC spectrum most likely comes from the secondary. 

\subsection{Lithium in the individual components of VLM binaries from the literature}
 
We made a search for published results to put our results for \dw{} and \dwr{} in context. The criteria employed to select the VLM binaries for this study are the following: total mass $<$\,150\,$M_\mathrm{Jup}$, error in total mass $<$\,10\%, spectral type of the components between M8.5$\pm$0.5 and T3.5$\pm$0.5\@, and individual dynamical mass of each component determined from orbital motion. The early spectral type limit at M8.5$\pm$0.5 corresponds to the lithium depletion boundary (LDB) at an age of about 150 Myr. At an age of 112$\pm$5 Myr the LDB has been determined at M7 in the Pleiades cluster by \citet{Dahm:2015}. The late spectral type limit corresponds to the latest dwarf with lithium detection, beyond which there might be significant chemical depletion of pure neutral lithium atoms due to the formation of Li-bearing molecules 
\citep{1999ApJ...519..793L, gharib21}. Even though it is predicted that the Li\,{\small{I}} resonance doublet could remain strong down to Teff\,=\,500\,K \citep{2002ApJ...577..986Burrows}, this feature has not yet been observed at spectral type later than T3.5$\pm$0.5 (this work), and thus we limit our study to objects not later than T3$\pm$1.0, corresponding to Teff\, about \,1100\,K. The relevant information available in the literature for the sample of seven VLM binaries that meet our criteria, together with that for \dw{} and \dwr{}, are summarized in Table \ref{tab_Phan2:LiMass}.  

Two of the binaries, namely Epsilon Indi\,Ba,b, and Luhman\,16\,AB, are the two closest substellar systems to the Sun, they have dynamic mass measurements 
\citep{Dupuy:2019aa,Lazorenko:2018aa}, and they have spatially resolved optical spectra of the individual components with sufficient quality to detect the Li\,{\small{I}} resonance doublet. It is indeed detected in both components of Luhman\,16\,AB \citep{faherty14a,2015A&A...581A..73L}, but not in the components of Epsilon Indi\,B 
\citep{2010AA...510A..99King}. Epsilon Indi\,Ba has a spectral type of T1$\pm$1, which is just at the edge of our range of study (considering the uncertainty), and we include it in the analysis. Epsilon Indi\,Bb has a spectral type of T6$\pm$1, outside our range, and we exclude it from the analysis. 

The other five binaries from the literature have published optical spectra for the spatially unresolved light of the components. In the previous section we estimated upper limits to the pEW of the Li\,{\small{I}} resonance doublet in each individual component of \dw{}. Here, we apply the same method to the four other binaries in our sample for which upper limits on the pEW of the Li\,{\small{I}} resonance doublet of the integrated light are available in the literature. 

The first one that we consider is GJ\,569\,B, which is composed of two M9 components  
with an observed magnitude difference of 0.7$\pm$0.2\,mag in the HST $F814W$-band filter 
\citep{2006AA...456..253Martin}. Since the two components have similar optical spectral type, we assume that that they also have similar $R-I$ color and that the flux ratio in the spectral region of the Li\,{\small{I}} feature is 1.9$\pm$0.4, which implies that the upper limit of pEW Li\,{\small{I}} =  0.05\,\AA{} from a Keck HIRES spectrum 
\citep{Zapatero-Osorio:2005aa} translates into an upper limit of 0.08\,\AA{} for the primary and 0.17\,\AA{} for the secondary. 
 
The second system is 2MASS\,J07003664$+$3157266AB which is composed of an L3$\pm$1 primary and an L6.5$\pm$1.5 secondary with a magnitude difference of 1.49$\pm$0.02 mag in the $J$-band \citep{Dupuy:2012fk}. Using the $R-J$\,=\,5.91$\pm$0.05 color of the L3 standard, and the average color $R-J$\,=\,6.2$\pm$0.2 mag of the L6 and L7 standards from \citet{2006PASP..118..659L}, we estimate a magnitude difference of 2.3$\pm$0.3 mag in the $R$-band, and a flux ratio of 8.3$\pm$2.7\@. The upper limit of pEW Li\,{\small{I}} of 0.3\,\AA{} in the combined light is from a spectrum obtained by \citet{2003PASP..115.1207Thor}. 
From these considerations, we obtain pEW Li\,{\small{I}} $<$\,0.3\,\AA{} for 2MASS J0700$+$3157\,A and pEW Li\,{\small{I}} $<$\,3.6\,\AA{} for 2MASS\,J0700$+$3157\,B\@. 
%5.98, 6.11, 6.08, 5.86, 5.92, 5.79, 5.30, 5.86, 5.53, 5.73, 6.21, 5.92 L2 -- L3.5 ... 6.30, 6.85, 6.02, 6.03, 6.09, 6.19, 6.08 L5 -- L7.5

The third binary is 2MASS\,J2132114$+$134148\,AB\@. The A component has a spectral type of L4.5$\pm$1.5, the B component has a spectral type of L8.5$\pm$1.5, and the magnitude difference is 0.85$\pm$0.04 mag in the J-band \citep{Dupuy:2012fk}. Using again the $R-J$ data of \citet{2006PASP..118..659L}, a flux ratio of 3.1$\pm$0.6 in the $R$-band is obtained. An optical spectrum of 2MASS\,J2132$+$1341\,AB was obtained 
by \citet{2007AJ....133..439Cruz} using the Gemini Multi-object spectrometer (GMOS) with a resolving power of R\,=\,1220\@. 
These authors do not note the presence of the Li\,{\small{I}} resonance doublet in their GMOS data. On the other hand, they note the presence of the Li\,{\small{I}} feature in a spectrum of the L5 dwarf 2MASS\,J0310140$-$275645, which has almost the same apparent near-infrared magnitudes as 2MASS\,J2132$+$1341\,AB, and it was observed with the same GMOS intrumental configuration. %\textcolor{red}{the most important here is the instrument and the set-up: is it the same resolution ?} J=14.86 L4 22.5pc
They provide a measurement of pEW Li\,{\small{I}}\,$=$\,10$\pm$1\,\AA{} for 2MASS\,J0310140$-$275645\@. Plots of the spectra of both 2MASS\,J0310140$-$275645 and 2MASS\,J2132$+$1341\,AB are available at http://database.bdnyc.org/query. We zoomed the plots on the spectral region from 667.0 nm to 687.0 nm, and we digitized them using the online tool WebPlotDigitizer. Using the IRAF task splot, we obtain a pEW Li\,{\small{I}} consistent with the published value for 2MASS\,J0310140$-$275645 and an upper limit for 2MASS\,J2132$+$1341\,AB of pEW Li\,{\small{I}}\,$<$\,0.8\,\AA{} from which we derive pEW Li\,{\small{I}}\,$<$\,1.1\,\AA{} for the A component and pEW Li\,{\small{I}}\,$<$\,3.8\,\AA{} for the B component.  

\begin{table*}
%\centering
\caption{Lithium and dynamical mass data for VLM binaries (total binary mass $< 150$\,$M_\mathrm{Jup}$).}
 \begin{tabular}{@{\hspace{0mm}}l c c c c c c@{\hspace{0mm}}}
 \hline
 \hline
Name  &  pEW(Li\,{\small{I}})$_1$  &  pEW(Li\,{\small{I}})$_2$ & M$_1$  & M$_2$  & Age & SpT   \cr
% \hline
      &    \AA{}                   & \AA{} & M$_\mathrm{Jup}$ & M$_\mathrm{Jup}$ & Myr &   \cr
 \hline
GJ\,569\,Ba,Bb       & $<$ 0.08 & $<$ 0.17 & $80^{+9}_{-8}$ & $58^{+7}_{-9}$  & 440$\pm$60 & M9,M9  \cr
Epsilon Indi\,Ba,Bb       & $<$ 0.10 & $<$ 0.10 & $68^{+0.9}_{-0.9}$ & $53.1^{+0.3}_{-0.3}$  & 4000$\pm$500 & T1,T6  \cr
2MASS\,J0700$+$3157\,AB       & $<$ 0.35 & $<$ 3.6 & $68^{+2.6}_{-2.6}$ & $73^{+2.9}_{-3}$  & 2000$\pm$700 & L3,L6.5  \cr
2MASS\,J2132$+$1341AB       &  $<$ 1.1 &  $<$ 3.8 & $68^{+4}_{-4}$ & $60^{+4}_{-4}$  & 1440$\pm$370 & L4.5,L8.5  \cr
 \dwr{}    &  $<$ 0.51 &  10.2$\pm$2.5 & $59^{+5}_{-5}$ & $41^{+4}_{-4}$  & 1110$\pm$220 & L4.5,T3.5  \cr
\dw{}       & $<$ 0.18 & $<$ 2.7 & $54^{+10}_{-9}$ & $54^{+6}_{-5}$  & 200$\pm$50 & M8.5,L4.5  \cr
Gl\,417\,BC       & 16.8$\pm$0.3 &  & $51.5^{+1.7}_{-1.8}$ & $47.7^{+1.9}_{-1.9}$  & 490$\pm$40 & L4.5,L6  \cr
SDSS\,J0423$-$414AB       & 11.4$\pm$0.7 & & $51.6^{+2.3}_{-2.5}$ & $31.8^{+1.5}_{-1.6}$  & 810$\pm$90 & L6.5,T2.5  \cr
Luhman\,16\,AB       & 8.2$\pm$1.0 & 8.4$\pm$1.5 & $33.5^{+0.3}_{-0.3}$ & $28.6^{+0.3}_{-0.3}$  & 1000$\pm$800 & L7.5,T0.5  \cr
 \hline
\label{tab_Phan2:LiMass}
\end{tabular}
\end{table*}
 
For the two brown dwarf binaries with lithium detections, namely Gl\,417\,BC and SDSS\,J0423$-$0414AB %\textcolor{red}{need references for each object here} 
we check whether or not it is reasonable that both components have preserved their initial lithium content. Gl\,417\,BC is composed of a L4.5$\pm$1 primary and a L6$\pm$1 secondary with a magnitude difference of 0.347$\pm$0.025 mag in the $K$-band \citep{Dupuy:2012fk}. According to the same authors, SDSS\,J0423$-$0414\,AB is composed of an L6.5 primary and a T2.5 secondary with a magnitude difference of 1.18$\pm$0.08 mag in the $K$-band. Using the same procedures as outlined above for binaries with components of similar spectral type, the flux ratios derived for the $R$-band are 1.9$\pm$0.1 and 27.8$\pm$0.9, respectively. 

\citet{2001AJ....121.3235Kirkpatrick} measured a pEW Li\,{\small{I}} of 11.5\,\AA{} for the combined-light spectrum Keck/LRIS of Gl\,417\,BC obtained with LRIS at Keck-I\@. They did not give an error bar, but from their Figure 2, we estimate $\pm$0.2\,\AA{}. We derive a pEW Li\,{\small{I}} of 16.8$\pm$0.3\,\AA{} for the A component. This is consistent with the upper envelope of pEW Li\,{\small{I}} measured in field L4.5 dwarfs (see Fig.\ \ref{fig:sptLi}). There could be a modest contribution of the B component to the Li\,{\small{I}} feature, but we cannot quantify it precisely with the data in hand. High spectral resolution spectroscopy might be able to resolve the two Li\,{\small{I}} features from the components in this system. 
%\textcolor{red}{other option is AO images to get a spectrum for each component ?}

\citet{Kirkpatrick:2008uk} measured a pEW Li\,{\small{I}} of 11\,\AA{} for the spatially unresolved spectrum of SDSS\,J0423$-$0414\,AB obtained with LRIS at Keck-I\@. They did not provide an error bar, but the spectrum shown in their Figure 16 is of high quality, and hence we assign an error bar of 0.5\,\AA{}, which is the upper limit they assign to their high quality objects without lithium detection. Using the flux ratio we obtain pEW Li\,{\small{I}} of 11.4$\pm$0.7\,\AA{} for the A component, which is also consistent with the upper envelope of Li measurements for L6.5 spectral class (see Fig.\ \ref{fig:sptLi}). The contribution of the B component is negligible because of the high flux ratio of 27.8$\pm$0.9 that we estimated previously. 
%\textcolor{red}{because it is a T2 dwarf and Li is in form of LiCl ? Need to justify your argument}  
%$R-K_S = 7.64\pm0.05$ mag. for the two L5 standards of (Liebert \& Gizis (2006)
%$R-K_S = 8.0\pm0.1$ mag. for the two L6 and L7 standards of (Liebert \& Gizis (2006)
%$R-K_S = 8.44\pm0.1$ mag. for the two L8 standard of (Liebert \& Gizis (2006)
%delta m R = 0.71 for Gl417BC

 \subsection{Unresolved pairs of brown dwarfs} 

It has been noted that some of the objects in our sample could be unresolved pairs of brown dwarfs because the evolutionary models cannot provide a consistent account of their age, luminosity and mass. %Simon et al.\ (2006) 
\citet{2006ApJ...644.1183Simon} suggested that GJ\,569Bb could be a pair of brown dwarfs with masses of about 40\,$M_\mathrm{Jup}$ each. \citet{Dupuy:2017aa} considered that 2MASS\,J0700$+$3157\,B could be composed of two equal mass brown dwarfs with mass of 36.7$^{+1.4}_{-1.5}$\,M$_\mathrm{Jup}$ each.  
\citet{2021MNRAS.500.5453S} proposed that \dwb{} is made up of two unresolved brown dwarfs with mass of 25\,$M_\mathrm{Jup}$ each. 

All these possible unresolved candidate pairs have individual masses well below those adopted in this work. Should it be demonstrated that any of them, or all of them, are indeed composed of two lower-mass objects, the non-detection of Li in them could be surprising. We strongly encourage additional observations of these three objects in particular to try to resolve them into their putative pairs, and try to get the lithium detections, spectral types, and individual masses. One possible solution is that the individual masses of the unresolved pairs are so low that their spectral types are cooler than T4, and they might be affected by significant weakening of the Li\,{\small{I}} resonance doublet due to Li locking into dust grains and molecules.

%which was proposed byto account for the flux ratio between \dwa{} and \dwb{} because objects of such low mass are far below the MMLB and ought to have preserved all of their initial lithium reservoir. 

\section{Ages and metallicities from stellar companions}

Three of the VLM binaries considered in this study are wide companions of stars and can be assumed they are %considered as 
members in hierarchical multiple systems sharing a common origin. We estimate here the ages and metallicities of these three stars, which can serve as reference in their respective systems. %to inform on the properties of the systems.

We inferred the stellar parameters of the three stars 
%age inferred from stellar evolution models 
using the Bayesian inference code of 
\citet{2016MNRAS.463.1400delBurgo,2018MNRAS.479.1953delBurgo}
%del Burgo \& Allende Prieto (2016, 2018) 
and incorporating the data from \textit{Gaia} DR2 \citep{Gaia-Collaboration:2018ae}. This statistical method %based on Bayes' theorem 
was applied to the PARSEC v1.2S library of stellar evolution models \citep{2012MNRAS.427..127Bressan}. It has been proven that its predicted masses for main-sequence stars are, on average, within 4\% of accuracy when compared with the dynamical masses of detached eclipsing binaries \citep[][]{2018MNRAS.479.1953delBurgo}.
%(Bressan et al. 2012; Chen et al. 2014, 2015; Tang et al. 2014). 
In this analysis, the code takes as input the absolute $G$ magnitude (obtained from the apparent magnitude and the parallax), $G_{BP}-G_{RP}$ colour, [Fe/H], and their uncertainties, returning theoretical predictions for a series of %other 
stellar parameters. We adopted the corrected Gaia photometry passbands ($G$, $G_{BP}$, and $G_{RP}$) from \citet{2018A&A...619A.180Maiz}.
%Ma\'\i z Apell\'aniz \& Weiler (2018). 
The values for the iron abundance [Fe/H] were averaged from those with given uncertainties for these stars in the PASTEL catalogue, which is based on high-resolution spectra \citep{2016A&A...591A.118Soub}.
%(Souriban et al. 2016)
%{\color{red} This reference together with others are in the latex file I sent}. 
%In the case of BD$+$162708, with a single value, we adopted %assumed 
%an uncertainty of 0.2 dex for [Fe/H]. 

\begin{table*}
  \caption{
Literature parameters of the primaries: spectral type ($SpT$)/luminosity class ($LC$), \textit{Gaia} photometry ($G$, $G_{BP}$, $G_{RP}$), parallax $\Pi$, and 
iron abundance [Fe/H].
References for the literature values are given in the main text. 
}
  \label{tab:input}
  \begin{tabular}{lcccccccc}
    \hline
    Name            & Alternative name &  $SpT/LC$  &       $G$                     & $G_{BP}$                    & $G_{RP}$                & $\Pi$                       & [Fe/H]                     \\
                          &                 &                    &      ($mag$)               & ($mag$)                       & ($mag$)                   & ($mas$)                  & (dex)                      \\
    \hline
 Gl 417             & HD 97334  & G2/?  & 6.2410 $\pm$ 0.0006 & 6.5882 $\pm$ 0.0028 & 5.802 $\pm$0.005    &  44.14 $\pm$ 0.04  &   0.07 $\pm$ 0.04   \\
 GJ 569 A          & BD$+$162708  & M3V  & 9.1256 $\pm$ 0.0008 & 10.350 $\pm$ 0.003   & 8.059 $\pm$0.003    & 100.68 $\pm$ 0.05 &   -0.15 $\pm$ 0.20  \\
 Epsilon Indi     &  HD 209100  & K5V  & 4.273 $\pm$ 0.005     & 4.9221 $\pm$ 0.0008 & 3.6386 $\pm$0.0018 & 274.80 $\pm$ 0.25 &   -0.13 $\pm$ 0.09  \\
\hline
\label{tab_Phan2:LitStars}
  \end{tabular}
 \end{table*}

 \begin{table*}
  \caption{
Inferred stellar parameters of the primaries:  iron abundance [Fe/H], radius ($R$), effective temperature ($T_\text{eff}$), mass ($M$), surface gravity ($\log g$), luminosity ($\log L$), 
age ($\tau$), and evolution phase. 
}
  \label{tab:prop}
  \begin{tabular}{lccccccc}
    \hline
    Name              &  $R$                             & $T_\text{eff}$       & $M$                         & $\log g$                   &  $\log L$                & $\tau$ & phase \\
                         &   (R$_{\sun}$)                & ($\,K$)                 & (M$_{\sun}$)           & ($g$: cm s$^{-2}$)  & ($L$:  L$_{\sun}$)  &  $Myr$      &           \\
    \hline
Gl 417            &   1.032 $\pm$ 0.009     &  5913 $\pm$ 18  &  1.063 $\pm$ 0.028  &  4.437 $\pm$ 0.017 & 0.070 $\pm$ 0.005 &  2600 $\pm$ 1180      &  MS \\
%\hspace{12pt}$\shortparallel$            &   1.026 $\pm$ 0.009     &  5927 $\pm$ 21  &  1.109 $\pm$ 0.027  &  4.460 $\pm$ 0.016 & 0.069 $\pm$ 0.004 &  41 $\pm$ 18      &  PMS \\
GJ 569 A         &   0.540 $\pm$ 0.010     &  3432 $\pm$ 42  &  0.44 $\pm$ 0.07      &  4.61 $\pm$ 0.10     & -1.438 $\pm$ 0.008 &  87$^{+153}_{-87}$   &  PMS \\
Epsilon Indi      &   0.7068 $\pm$ 0.0028 &  4676 $\pm$ 8    &  0.730 $\pm$ 0.014   & 4.603 $\pm$ 0.010 & -0.667 $\pm$ 0.004 &  9266$\pm$2114       &  MS \\
\hline
\label{tab_Phan2:ResStars}
  \end{tabular}
 \end{table*}

Gl 417 A: Our Bayesian analysis provides %two different solutions, one %on the Main-Sequence (MS) with 
a most likely mean age of 2.60$\pm$1.18 Gyr, on the Main-Sequence (MS).
%, and, if imposing (as a prior {\bf***Leave it like this***}) to be on the Pre-Main-Sequence (PMS), a mean age of 41$\pm$18 Myr.
\citet{2001AJ....121.3235Kirkpatrick} estimated an age in the range from 80 to 300 Myr, %which is 
based on several indicators such as the X-ray luminosity (log(L$_X$/L$_{bol}$)= 4.55), the CaII emission (log=-4.40), and the rotational period (7.6 days). However, \citet{Dupuy:2019aa} derived a gyrochronology age of 750$^{+150}_{-120}$ Myr and found the estimate of \citet{2001AJ....121.3235Kirkpatrick} inconsistent.
The TESS light curve also yields a rotational period of about 7.5 days, which suggests an age similar to the Praesepe open cluster (670 Myr) according to the calibration provided by \citet{2019ApJ...879...49Curtis}. The age inferred from the substellar evolution models is 0.49$\pm$0.04 Gyr \citep{Dupuy:2017aa}, which is %more 
in good %reasonable 
agreement %consistent 
with our estimate %the MS age 
(within 1.8$\sigma$).
%, than with the PMS age (25$\sigma$).  

GJ 569 A: Our analysis places it on the pre-MS phase, with an upper limit for the mean age $<$240 Myr. This is somewhat lower, but not far from the value of 300 Myr from its kinematics, which places it as a member of the Ursa Major moving group \citep{2001ApJ...554L..67Kenworth}. The age inferred from the substellar evolution models is 440$\pm$60 Myr \citep{Dupuy:2017aa}, which is somewhat older than the PMS age that we obtain for the stellar primary, but within 1.75$\sigma$.  
%We obtain an iron abundance of [Fe/H]\,=\,$-$0.15$\pm$0.19 dex. 
%\textcolor{red}{no TESS light curve for it but the association to a YMG is interesting. Is Kenworthy the only reference about the age of Gl569 ?}
%\textcolor{blue}{I provided Eduardo all the references I found in the literature, but it may be others.}

Epsilon Indi A: We infer a mean age of 9$\pm$2 Gyr (on the MS). This is consistent with the value of $>$9.87 Gyr derived from the kinematics \citep{2015AJ....149..131Eker}, but greater than the range of 3.7$-$5.7 Gyr given by \citet{2019ApJS..242...25Feng}, which is based on the same rotation-age relationship of \citet{2015AJ....149..131Eker}.  
%We obtain an iron abundance of [Fe/H]\,=\,0.04$\pm$0.04 dex. 
%In this case, although we fixed the metallicity to be [Fe/H]=-0.13$\pm$0.09, our Bayesian approach points at a solar metallicity ([Fe/H]=0.04$\pm$0.04). 
The TESS light curve indicates a rotation period longer than 18 days, which would imply an age older than 1 Gyr \citep{2019ApJ...879...49Curtis}. The age from the substellar evolution models is 4.0$\pm$0.5 Gyr \citep{Dieterich:2018aa}, which a factor of two younger than our Bayesian estimate, comparing mean values, but only within 2$\sigma$ %level 
of discrepancy in view of the uncertainties.

For all the three stellar companions, our Bayesian analysis and those obtained by other authors for the brown dwarfs are in agreement within 2$\sigma$ uncertainty level. 
%Because:
%Epsilon Indi: 4+0.5+0.5=5; 5=9-2-2
%Gl417: 0.49+0.04+0.04=0.57; 0.24=2.60-1.18-1.18
%GJ569: 440-60-60=320; 393=87+153+153
%The maximum discrepancy between the ages obtained from our Bayesian analysis of the stellar companions and those obtained by other authors for the brown dwarfs is for the GJ 417 system; the stellar mean age obtained from our Bayesian approach is about a factor of five older than the substellar one, but within 1.8$\sigma$ considering the uncertainties. For the other two systems the coincidence between the stellar and substellar ages is better than a factor of three and 2 $\sigma$ level. For all the systems considered we find that the discrepancies between the ages obtained from the stellar and substellar evolutionary models do not exceed the 2$\sigma$ (not 3, but 2) uncertainty derived from the Bayesian analysis. 
% ({\textcolor{red}{***discuss it in terms of sigma***}}). 
We note that our conclusions on the location of the MMLB does not depend on the precise age adopted for the systems under study, as long as they are not extremely young ($<$150 Myr). If any of the systems under study would be younger than about 150 Myr, the MMLB could be shifted to higher massess because the brown dwarfs may not have had enough time to complete their Li burning.  
%In future work we plan to study whether or not the discrepancies in age estimates obtained from different methods are significant. %{\texcolor{red}{We could provide the differences if any.}}
%\textcolor{red}{where do you get the ages of the BDs from ? You have no reference and no numbers so it is impossible to check the factors that you give. The masses of the BD around Epsilon Indi certainly give an age older than 1 Gyr if I recall but the upper limit is hard to define.}

The iron abundances %obtained from our analysis 
of the three stars are similar to the Sun (within about 50\%), providing support to our assumption of solar metallicity for the comparison of the observed MMLB with the theoretical MMLB, which is discussed in the next section.

\section{The minimum mass for thermonuclear lithium burning in brown dwarfs}

\begin{figure}
\centering
\includegraphics[width=\linewidth]{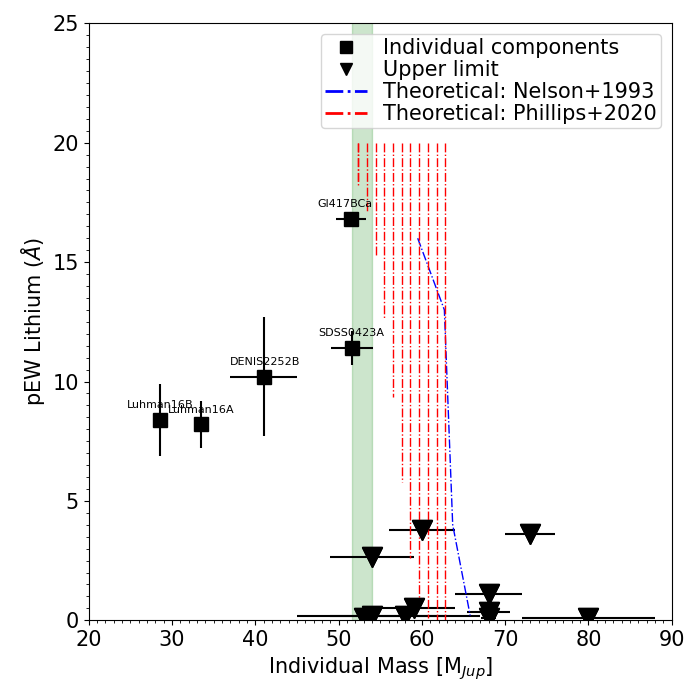}
\caption{Individual dynamical mass versus pEW (Li\,{\small{I}}) for components of VLM binary systems (total binary mass $<$\,150\,M$_\mathrm{Jup}$). Measurements are shown with black squares while upper limits are displayed as triangles. Blue dashed lines represent the calculations of Li burning by 
\citet{1993ApJ...413..364Nelson}. 
Red dashed lines represent the isomasses between 51 and 63 Jupiter masses from the models by  \citet{2020AA...637A..38Ph} for ages between 100 Myr and 5000 Myr. Theoretical prediction of Li depletion factors have been converted into pEW Li\,{\small{I}} resonance doublet using the prescription provided by \citet{Martin:2018aa}, assuming an upper limit on the pEW of 20\,\AA{}. The green vertical band shows the sharp region between Li-bearing and Li-free brown dwarfs. 
}
\label{fig:massLi}
\end{figure}

Detailed theoretical calculations of the minimum mass for thermonuclear lithium burning (MMLB) have been carried out by several groups using different numerical codes. \citet{1991MmSAI..62..171Pozio} found that a 63 $M_\mathrm{Jup}$ model with solar composition could destroy 90\% of its initial Li content. 
\citet{1993ApJ...413..364Nelson} obtained slightly less Li depletion, i.e.\ their model with 65 $M_\mathrm{Jup}$ destroyed 90\% of lithium.\citet{1996ApJ...459L..91Chabrier} found that their model with $63$\,$M_\mathrm{Jup}$ was able to destroy 99\% of lithium. The very recent models with updated equation of state and opacities 
\citep{2020AA...637A..38Ph}
destroy 90\% of lithium at a mass of $58$\,$M_\mathrm{Jup}$. In this work we adopt as the definition of the MMLB the mass at which the permanent imprint of thermonuclear lithium fusion leaves an amount of 10\% of the original Li abundance in the photosphere of the brown dwarf.  
The calculations of \citet{1993ApJ...413..364Nelson} and \citet{2020AA...637A..38Ph} bracket the range of theoretical MMLB values found in the literature, which we summarize as MMLB$_{th}$ = 63$\pm$5$\,M_\mathrm{Jup}$. 
%(MMLB$_{observed}$=52.5$\pm$1.5~$M_\mathrm{Jup}$) . 
%All published models of brown dwarf evolution predict that the LMTLB is located at $>60$\,$M_\mathrm{Jup}$, meaning that at least 10\% of the initial lithium content is expected to be preserved forever in objects of lower mass. 

The availability of individual dynamical masses, effective temperatures, bolometric luminosities and lithium data for the components of very low-mass (VLM) binary systems makes it possible to test observationally the theoretical predictions for the hydrogen and lithium burning minimum masses. An early attempt to carry out such a test was performed by \citet{Zapatero-Osorio:2005aa} using only GJ 569B, but the uncertainties were to high to draw strong conclusions. Based on a sample of 13 VLM  binaries, the minimum mass for hydrogen burning (MMHB) was estimated at $70$\,$M_\mathrm{Jup}$ and the MMLB at $60$\,$M_\mathrm{Jup}$ by \citet{Dupuy:2017aa}. They noted that their results were in agreement with the models of \citet{Baraffe:2015aa}, but not with those by   \citet{1997ApJ...491..856Burrows}. 

A comparison between the pEW versus dynamical mass data and two different sets of evolutionary models is illustrated in Fig.\ \ref{fig:massLi}. The data and the models coincide on the fact that there is a sharp mass boundary between lithium depletion versus preservation in brown dwarfs. However, the data suggest that the boundary is displaced to lower mass than what the models predict, as can be appreciated by eye from the location of the boundary between the objects with and without lithium, depicted as a green band in Fig.\ \ref{fig:massLi}, and the location of the theoretical curves. Random uncertainties in the dynamical mass measurements could make the objects scatter around the true MMLB, making it appear as if some objects with lithium could have a mass higher than other objects without lithium. Other factors such as a spread in age or metallicity could also make the MMLB wider. However, that is not observed, suggesting that mass is by far the single most dominant factor. 

\subsection{Lithium abundances, lithium depletion factors and the observed mass boundary}

We used the pEW values listed in Table \ref{tab_Phan2:LiMass} and the basic parameters from the literature \citet{2015A&A...581A..73L, Dupuy:2017aa, Dupuy:2019aa}, to calculate the Li abundances or upper limits for 14 individual components of VLM binaries using the same procedures as in our previous work on L-type members in the Coma Ber and Hyades open clusters \citep{Martin:2018aa, martin20a}. The resulting abundances and depletion factors are summarized in Table \ref{tab_Phan2:LiAbundances}. For each object, we also estimated the Li depletion factor predicted by the models of \citet{2020AA...637A..38Ph} using the age and mass estimates given in Table \ref{tab_Phan2:LiMass}. We find a good agreement between the Li depletion factors estimated from the observations and those predicted by the models, except for the two components of \dw\, which are both more depleted than the theoretical expectation. Therefore, we note that this binary system is particularly critical to test the validity of the models.

\begin{table*}
%\centering
\caption{Lithium abundances and depletion factors for individual components of VLM binaries.}
 \begin{tabular}{@{\hspace{0mm}}l c c c c c c c@{\hspace{0mm}}}
 \hline
 \hline
Name  &  log (Lbol/L$_\odot$) & Teff & log g & pEW(Li\,{\small{I}})$_{obs}$/  &  logN(Li) & Li depletion  & Li depletion \cr
% \hline
      &  & (K) & & pEW(Li\,{\small{I}})$_{max}$ &  & (observed) & (predicted) \cr
 \hline
GJ\,569\,Ba      & -3.44$\pm$0.04 & 2420$\pm$40 & 5.2 & $<$0.04 & $<$0.5  & $>$99.5\% & 100\%  \cr
GJ\,569\,Bb      & -3.67$\pm$0.03 & 2170$\pm$50 & 5.2 & $<$0.11 & $<$1.3  & $>$95\%   & 1-100\% \cr
Epsilon Indi\,Ba      & -4.70$\pm$0.02 & 1310$\pm$10 & 5.4 & $<$0.01 & $<$0.1 & $>$99.9\% & 100\% \cr
2MASS\,J0700$+$3157\,A  & -3.95$\pm$0.04 & 1860$\pm$40 & 5.4 & $<$0.05 & $<$0.6 & $>$99.5\% & 100\% \cr
2MASS\,J0700$+$3157\,B  & -4.45$\pm$0.04 & 1430$\pm$100 & 5.4 & $<$0.24 & $<$2.0 & $>$90\% & 100\% \cr
2MASS\,J2132$+$1341A   & -4.22$\pm$0.05 & $1660^{+50}_{-40}$ & 5.3 &  $<$0.06 &  $<$0.8 & $>$99.5\% & 100\% \cr
2MASS\,J2132$+$1341B   & $-4.50^{+0.05}_{-0.04}$ & $1400^{+30}_{-40}$ & 5.3 & $<$0.42 & $<$2.5 & $>$80\% & 35-100\% \cr
\dwra{}   & $-4.26^{+0.05}_{-0.04}$ & 1590$\pm$50 &  5.3 &  $<$0.05 &  $<$0.6 &  $>$99.5\% &  85-100\%  \cr
\dwrb{}   & $-4.76^{+0.08}_{-0.07}$ & $1210^{+50}_{-40}$ &   5.2 & 1.0$\pm$0.1 & 3.3$\pm$0.1 & 0-10\% & 0\%  \cr
\dwa{}                & -3.3$\pm$0.1 &  2400$\pm$100 & 5.1 & $<$0.10 & $<$1.3 & $>$99\% & 0-93\% \cr
\dwb{}                 & -4.1$\pm$0.1 &  1400$\pm$100 & 5.1 & $<$0.16 & $<$1.7 & $>$97.5\% & 0-36\% \cr
Gl\,417\,B             & -4.13$\pm$0.03 & $1639^{+29}_{-31}$ & 5.1 & 1.0$\pm$0.1 & 3.3$\pm$0.1 & 0-10\% & 1-11\% \cr
SDSS\,J0423$-$414A     & -4.41$\pm$0.04 & $1430^{+30}_{-40}$ & 5.2 & 0.8$\pm$0.2 & 3.1$\pm$0.3 & 0-70\% &  1-14\% \cr
Luhman\,16\,A         & -4.68$\pm$0.08 & $1305^{+180}_{-135}$ & 5.0 & 0.8$\pm$0.2 & 3.1$\pm$0.3 & 0-70\% & 0\% \cr
Luhman\,16\,B         & -4.71$\pm$0.10 & $1190\pm60$ & 4.9 & 1.0$\pm$0.1 & 3.3$\pm$0.1 & 0-10\% & 0\% \cr
 \hline
\label{tab_Phan2:LiAbundances}
\end{tabular}
\end{table*}
 
 We performed curve fitting for our compilation of Li abundance versus individual dynamical mass. 
 To explore the constraints obtained from the data on the location of the MMLB, we chose a natural exponential decay function $a \times exp(-b(x-c))$ and excluded the two points corresponding to the binary components of the system Luhman 16AB as well as Reid1 B  because they are well below the MMLB boundary. We neither employ Epsilon Indi Bb in the analysis.
The \textit{curve\_fit} function of the Open Source  Python\footnote{https://docs.python.org} scientific library \textit{SciPy} \citep{virtanen20} was used to derive the free parameters $a$, $b$, and $c$. 
A Monte Carlo simulation was carried out %allowed us 
to derive the uncertainties. %parameters. 
We generated 10\,000 distributions of points and applied our fitting approach %the explained fitting procedure, 
rejecting bad fits before extracting the statistics. 
Assuming the original pairs and their uncertainties %correspond to normal distributions, 
adjust to Gaussian distributions, 
we employed the \textit{NumPy}\footnote{https://numpy.org/} function \textit{random.normal} to generate suitable %random 
samples. 
For the upper limits of Li abundance, we used the \textit{NumPy} function \textit{random.random}. The exponential that fits the original distribution has a very steep decay. For $c$ we adopted the value that better fits this distribution and increased its uncertainty in view of the shift with respect to the most likely value of $c$ that emerges from the Monte Carlo simulation. Our approach yields $c= 51.48^{+0.22}_{-4.00}$ M$_{Jup}$. 

\begin{figure}
\centering
\includegraphics[width=\linewidth]{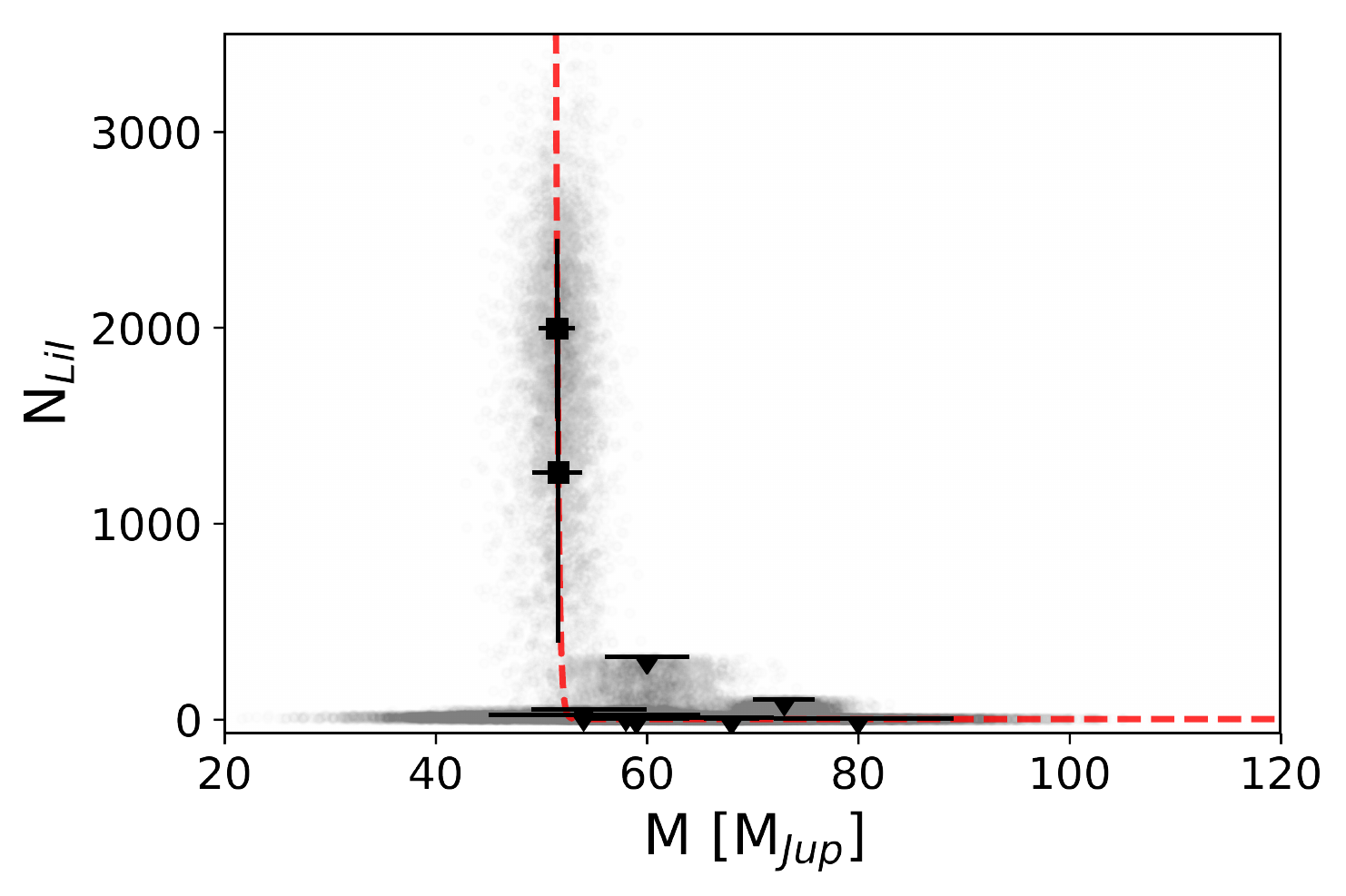}
\caption{Number of Li atoms in the atmosphere of brown dwarfs versus their individual dynamical mass for our sample of components of VLM binary systems. Measurements are shown with black squares and error bars, while upper limits are displayed as down-pointed triangles with error bars on the masses. A probabilistic cloud of simulated data with a Monte-Carlo code is associated to each datapoint. The best fit %to the Monte-Carlo simulation using an 
using a natural exponential law is shown with a red curve. From this fit, we obtain MMLB$_{obs}$\,=\, 51.48$^{+0.22}_{-4.00}$~$M_\mathrm{Jup}$.  This figure was made with the Matplotlib package \citep{hunter2007}}.
\label{fig:massLi-sim}
\end{figure}

An exponential law is a well-motivated approximation based on the estimate that thermonuclear Li burning is a extremely strong function of the maximum central temperature attained in the interior of brown dwarfs \citep{1997ApJ...482..442Bildsten}.
%(Bildsten et al. 1997). 
Besides, the observed data distribution supports it. In conclusion, we obtain
MMLB$_{obs}$= 51.48$^{+0.22}_{-4.00}$~$M_\mathrm{Jup}$. 
The result and scaffolding of this fitting exercise are illustrated in Fig.\ \ref{fig:massLi-sim}.

\section{Final remarks and future prospects}

The main goal of this work is to set observational constraints on the MMLB for solar metallicity and intermediate age brown dwarfs, and to test the validity of evolutionary models. We have considered a total of 15 individual components that belong to nine different binary systems, six of them are in common with the study of \citet{Dupuy:2017aa}. The new binaries considered in this work are GJ\,569\,Ba,Bb, Epsilon Indi Ba,Bb, and \dw{}, and we have provided additional data for the binary \dwr{}.  

From the study of this sample of 15 objects with dynamical mass and lithium data, we have determined a value of MMLB$_{obs}$=51.48$^{+0.22}_{-4.00}$ $M_\mathrm{Jup}$, which is lower than all of the available theoretical determinations of the MMLB location by the following amount: MMLB$_{th} - MMLB_{obs} = \delta$MMLB= 12.0$^{+9.0}_{-4.8}$~$M_\mathrm{Jup}$. This difference is minimized, but not cancelled for the models computed by \citet{2020AA...637A..38Ph}.
There seems to be a historical trend that when the evolutionary models get updated their predicted MMLB diminishes slightly. Further updates may bring the observational and theoretical MMLB to a complete agreement. 

Uncertainties in the models of lithium depletion for very low mass stars and brown dwarfs have been discussed in detail by \citet{2004ApJ...604..272B}, 
%Burke, Pinsonneault \& Sills (2004), 
although they neither consider masses below $68$\,$M_\mathrm{Jup}$, nor ages older than 200 Myr. Sources of theoretical uncertainty considered by those authors include the equation of state, the treatment of convective transport, the opacity data, the outer (atmospheric) and inner boundary conditions, and the effects of rotation. Another point of interest for the theoretical determination of the MMLB are the effects of dark matter accretion in the evolution of brown dwarfs, which have been suggested to be of potential relevance for the location of the H-burning minimum mass 
%for the evolution of stars with masses between 0.9 and $5$\,$M_\mathrm{Sun}$ \citep{zentner11,2021MNRAS.503.5611Raen}
\citep{zentner11}. In view of the increasing precision on the determination of the MMLB from BD binaries, it would be worthwhile to explore the effects of different dark matter candidates in the lithium burning of fully convective stars and degenerate brown dwarfs, particularly in environments where condensations of dark matter may be expected, such as the Galactic center. 
%However, for solar metallicity brown dwarfs it seems that dark matter is not required to explain the observed MMLB location.  

From the observational side, it is clear that more work is worthwhile to increase the accuracy of dynamical masses, and to expand the sample of brown dwarf binaries with dynamical masses and lithium abundance determination. With a larger sample it should be possible to increase the statistical significance of the observational constraints on the MMLB, and to check on the effects of age, metallicity, rotation, and other factors that could play a role in shaping the imprint of lithium burning in brown dwarf evolution. In future work, we plan to present new observations of lithium in brown dwarf binaries aimed at increasing the constraints on the determination of the MMLB as a function of age, mass, metallicity, rotation, and possible accretion and accumulation of dark matter in environments close to the Galactic center. 

Besides testing substellar evolutionary models, the determination of the observational MMLB can provide a useful benchmark to understand the properties of transiting brown dwarfs for which dynamical mass determinations can also be made. There is a growing number of these systems being discovered thanks to several transit surveys (e.g., \citet{2021arXiv210508574Acton}).
%(e.g., Acton et al.\ 2021). 
Comparison of their present mass, presence of lithium, stellar age, and the observational MMLB can set interesting constraints on their mass loss rate. 

\section*{Acknowledgements}
This research has been supported by the Spanish Ministry of Economy and Competitiveness (MINECO) and the Fondo Europeo de Desarrollo Regional (FEDER) under grants PID2019-109522GB-C53 and PID2019-107061GB-C66\@. This research made use of the databases at the Centre de Donn\'ees astronomiques de Strasbourg (\url{http://cds.u-strasbg.fr});
NASA's Astrophysics Data System Service (\url{http://adsabs.harvard.edu/abstract\_service.html}), and the paper repositories at arXiv. 
This research has made use of the SIMBAD database, operated at CDS, Strasbourg, France and NASA's Astrophysics Data System.
This research has made use of the VizieR catalogue access tool, CDS, Strasbourg, France. The original description of the VizieR service was published in A\&AS 143, 23.
This work has made use of data from the European Space Agency (ESA) mission {\it Gaia} (\url{https://www.cosmos.esa.int/gaia}), 
processed by the {\it Gaia} Data Processing and Analysis Consortium (DPAC, \url{https://www.cosmos.esa.int/web/gaia/dpac/consortium}). 
Funding for the DPAC has been provided by national institutions, in particular the institutions participating in the {\it Gaia} Multilateral Agreement. This research made use of Matplotlib and Scipy, two community-developed core Python packages. 
We thank Gilles Chabrier and Isabelle Baraffe for kindly providing a digitized version of the calculations that are used in Fig.\ \ref{fig:massLi} and for comments on the manuscript. We are grateful for the helpful referee review provided by John Gizis, and we also thank Trent Dupuy, Yakiv Pavlenko, Rafael Rebolo, Victor Sanchez Bejar and Maria Rosa Zapatero Osorio for reviewing the manuscript and providing comments. Jun-Yan Zhang assisted with digitizing the spectra from the BDNYC database. 

\section*{Data availability}
The data underlying this article will be placed in the GTC archive.

\bibliographystyle{mnras}
\bibliography{biblionew3}

\bsp	
\label{lastpage}
\end{document}